\documentclass[a4paper,11pt]{article}
\usepackage{jinstpub} 

\usepackage{graphicx}
\usepackage{subcaption}
\usepackage{caption}

\title{Laser diagnostics for negative ion source optimization: insights from SPIDER at the ITER Neutral Beam Test Facility}

\author[1,3]{R. Agnello}

 \affiliation[1]{{\'{E}cole Polytechnique Fédérale de Lausanne (EPFL), Swiss Plasma Center (SPC)},
             {CH-1015 Lausanne},
             {Switzerland}}

\author[2]{M. Barbisan}
\author[2]{R. Pasqualotto}
\author[4,3]{B. Pouradier-Duteil}
\author[3]{E. Sartori}
\author[3]{A. Tiso}
\author[3]{B. Zaniol}

\affiliation[2]{{Istituto per la Scienza e la Tecnologia dei Plasmi},
                {Corso Stati Uniti 4}, 
                {Padova},
                {35127}, 
                {Italy}}

\affiliation[3]{{Consorzio RFX},
{Corso Stati Uniti 4},
{Padova},
{35127},
{Italy}}

\affiliation[4]{{Commissariat à l'\'{E}nergie Atomique (CEA)},
                {IRFM}, 
                {St. Paul lez Durance},
                {F-13108}, 
                {France}}

\abstract{The ITER Heating Neutral Beams (HNBs) require large, high-energy H/D atom beams (285/330 A/m² extracted current density, and 1/0.87 MeV acceleration energy, respectively for H and D). To address the associated challenges, the SPIDER negative ion RF beam source at the Neutral Beam Test Facility (NBTF) in Padova (Italy) serves as a full-scale source prototype with a 100 kV triode accelerator, for design validation and performance verification. SPIDER is equipped with two advanced laser diagnostics to monitor key plasma parameters; Cavity Ring-Down Spectroscopy (CRDS) is used to measure H$^-$\slash D$^-$ ion densities, while Laser Absorption Spectroscopy (LAS) tracks caesium neutral density in the source. These measurements are essential for optimizing negative ion production and meeting ITER source targets. We present diagnostic upgrade details, recent experimental results, and correlations with other machine parameters. Since CRDS relies on a single 4.637-meter-long optical cavity, the longest used in such sources, it has demonstrated sensitivity to alignment. Based on recent experimental experience, structural improvements are being implemented to enhance both stability and measurement reliability. LAS has mainly been employed as a tool to monitor the caesium conditioning status of SPIDER. Additionally, due to a distributed measurement over four lines of sight, LAS has proven effective in monitoring the caesium distribution within the source. 
This work demonstrates the essential role of laser diagnostics in developing ITER-relevant plasma sources and informs ongoing efforts to improve measurement accuracy in challenging environments.}

\keywords{Negative Ion Sources, plasma spectroscopy}

\begin{document}
\maketitle
\flushbottom

\section{Introduction}
\label{sec:intro}

In ITER, at least two Heating Neutral Beam (HNB) injectors are envisaged, supplying each about 17 MW power with a current of 40 A of deuterium and an energy of 1 MeV for a duration of about 1h \cite{hemsworth2017}. 
To achieve these demanding parameters, the Neutral Beam Test Facility (NBTF), is operating to study the physics and the technology of high power neutral beam injectors (NBIs) for fusion. NBTF comprises the 1 MeV, 17 MW prototype injector, MITICA, which is under construction, and the full-scale source SPIDER, equipped with a 100 keV accelerator, and operational since 2018 \cite{marcuzzi2023}. The volume where the plasma expands in SPIDER has dimensions of 24 cm $\times$ 89 cm $\times$ 179 cm, and the plasma discharges are ignited by eight RF antennas operating at a frequency of 1 MHz.
SPIDER's main goals are maximizing the extracted negative ion current while minimizing the co-extracted electron current, optimizing the plasma homogeneity, the caesium (Cs) evaporation, and, in general, investigating the physics of large negative ion beams.
To obtain a full picture of these phenomena, SPIDER is equipped with a large suite of plasma diagnostics including laser-based absorption techniques such as the Cavity-Ring Down Spectroscopy (CRDS) and the Laser Absorption Spectroscopy (LAS) \cite{pasqualotto2012}. These are common tools used in negative ions sources \cite{quandt1999,tsumori2021,fantz2011}, such as in ELISE, the half-size ITER negative ion source prototype for the NBIs, used to study the acceleration of large negative ion beams up to 60 keV \cite{heinemann}. CRDS and LAS are employed to measure the negative ion and neutral Cs density, respectively, in the vicinity of the plasma-facing grids. Both diagnostics provide direct, calibration-free measurements, although they are integrated along the line of sight (LOS). Other approaches are also available; one of them, discussed in the following paragraph, is Optical Emission Spectroscopy (OES), specifically the line ratio method \cite{fantz2006}.
Another but invasive technique consists in the analysis of the ion branch of Langmuir probe I-V characteristics, which has been applied in SPIDER \cite{poggi2023}, even though affected by large uncertainty.
A laser photodetachment diagnostic, capable of providing spatially resolved $H^{-}$ measurements, has been used in similar sources \cite{tsumori2021} and is under consideration for SPIDER, though its implementation is challenging due to limited accessibility and the invasive nature of the technique.
Fluid modeling including $H^-$ production is currently under development \cite{zagorski2025}, even though reliable estimates of $H^{-}$ density depend on precise knowledge of the ion dynamics near the PG, particularly Cs coverage and the fluxes of ionic and neutral species such as H atoms and positive ions ($H^{+}$, $H_2^{+}$, $H_3^{+}$). 
As evaporated caesium is the precursor for negative ion formation at the surface of the ion source, knowledge of its evaporation rate and volumetric density is crucial to assess beam performance. LAS is routinely used in experiments as it provides an indirect method to monitor the caesiation of the source.
Additionally, the Cs density obtained from LAS can be used as a constraint in collisional–radiative codes, thereby reducing the uncertainty in the derived plasma parameters \cite{pouradier2024}.
However, measuring only Cs density in the source is not sufficient to fully characterize the caesiation status. This is mainly because understanding the behaviour of caesium in the source also requires information about its surface deposition and its transport within the plasma volume, a complex process influenced by plasma density and temperature, source duty cycle, magnetic field topology, and the initial distribution of caesium.\\
In previous studies, we characterized the $H^{-}$ and Cs densities in SPIDER by varying machine parameters in dedicated experimental scans \cite{barbisan2022_Csfree,barbisan2021_CRDS,barbisan2023_SOFT} and extracting isolated beamlets. These scans provided a general overview of source behaviour. However, further insights on the plasma conditions could be obtained by correlating $H^{-}$ and Cs densities with additional diagnostics such as OES. Moreover, SPIDER is now operated by extracting several hundreds of beamlets, a configuration known to affect plasma conditions \cite{sartori2023}, thus representing a new operational regime that deserves new investigations.
This study focuses on the time evolution of negative ion and caesium densities during source operation, as observed in recent experimental campaigns, including trends from OES.


\section{Laser diagnostics in SPIDER}

\begin{figure}
	\centering
	\includegraphics[width=.6\columnwidth]{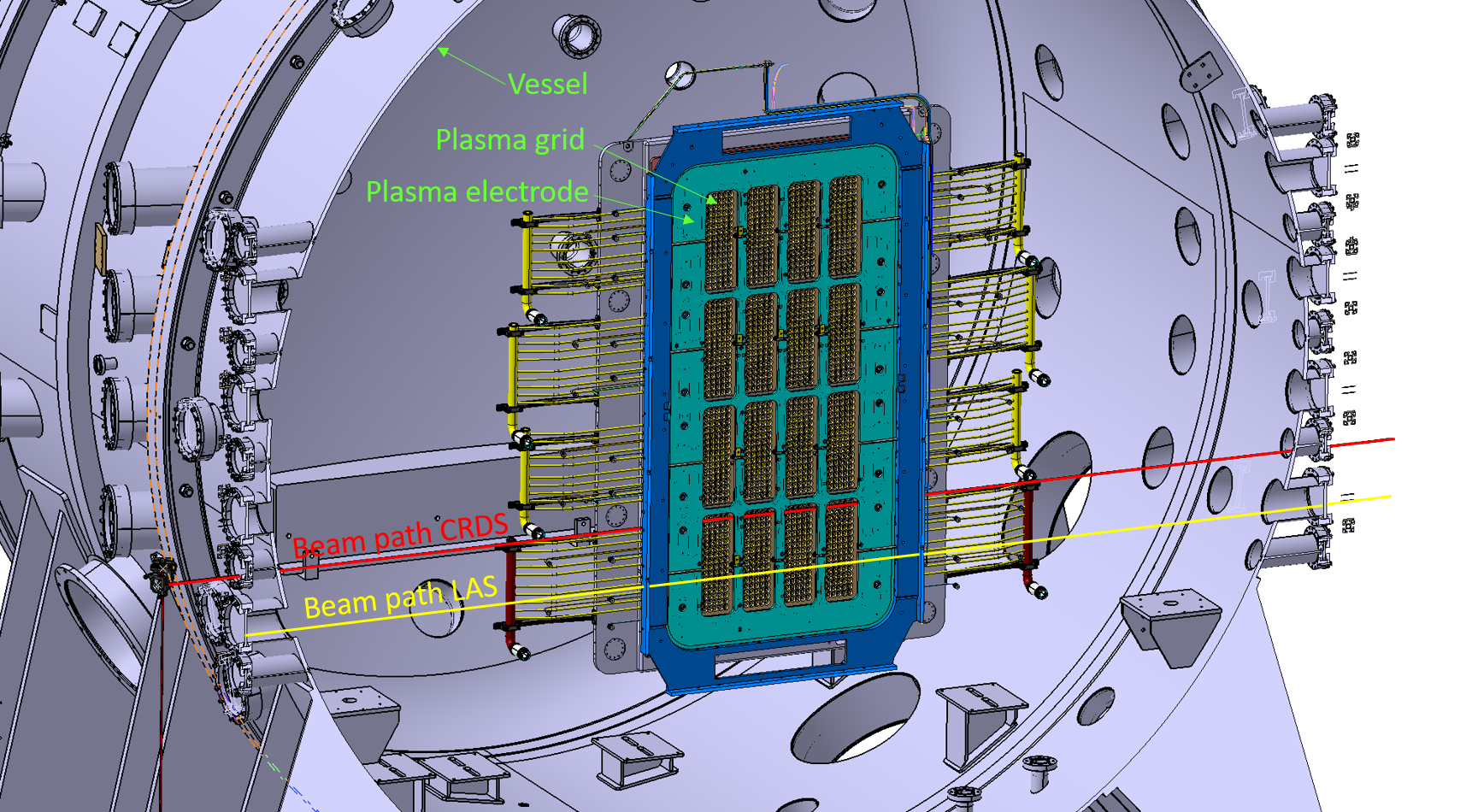}
	\caption{View of the laser beam path for CRDS and LAS in the context of the complete source and chamber.}
	\label{fig:LAS_CRDS_laser_path}
\end{figure}

Since negative ion production processes occur predominantly at surfaces, the LAS line of sight (LOS) is arranged to pass just above the bias plate (BP), whereas the LOS of CRDS passes between the plasma grid (PG) and the BP, as shown in Fig.\,\ref{fig:LAS_CRDS_laser_path}. From an operational standpoint, the alignment is particularly challenging, as the laser beam must pass through multiple 1 cm diameter apertures, while the alignment is performed outside the vessel wall, 1.830 m away from the source apertures.
We provide below a brief recall of how these techniques work on SPIDER.

\subsection{Cavity Ring-Down Spectroscopy}

CRDS in SPIDER relies on an optical cavity (with reflectivity $R = 99.994\%$) in which a 1064 nm laser pulse, with a duration of 5 ns is trapped.
Although the laser pulse has an energy of 150 mJ or less, only a small fraction, of the order of a few $\mu W$, actually enters the cavity, due to the very low transmissivity of the mirror coating.
The cavity mirrors are plano-convex, with a diameter of 25 mm and a radius of curvature of 6 m, ensuring cavity stability. The laser source is located in a concrete room outside the SPIDER bioshield to protect it from neutron and gamma radiation, and the beam is guided to the optical cavity through a series of mirrors.
The laser was operated at a repetition rate of 5 Hz, as set during the experiment, although the flash lamps would allow a maximum repetition rate of 10 Hz.
The difference in decay time with and without plasma is used to estimate the average H$^-$ density along the line of sight (LOS) of the measurement.
The negative ion density $n_{H^-}$ is then estimated as:
\begin{equation}
    n_{H^-}=\frac{L}{\sigma c d}\Big( \frac{1}{\tau}-\frac{1}{\tau_0} \Big)
\end{equation}
where $L$ is the cavity length, $\sigma$ the photodetachment cross section, which is equal to $3.5 \cdot 10^{-17}\,$ cm$^2$ at 1064 nm \cite{barnett1977}, $c$ the speed of light, $d$ the effective plasma length, $\tau$ and $\tau_0$ are respectively the decay time with and without the plasma discharge. 
\begin{figure}[htbp]
    \centering
    \begin{subfigure}[b]{0.32\textwidth}
        \centering
        \includegraphics[width=\textwidth]{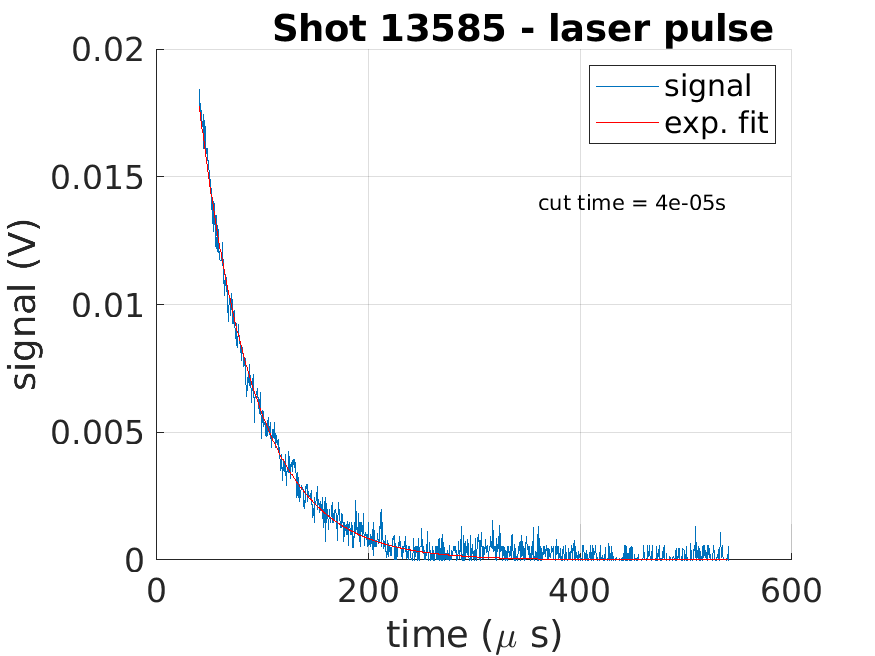}
        \caption{}
    \end{subfigure}
    \hfill
    \begin{subfigure}[b]{0.32\textwidth}
        \centering
        \includegraphics[width=\textwidth]{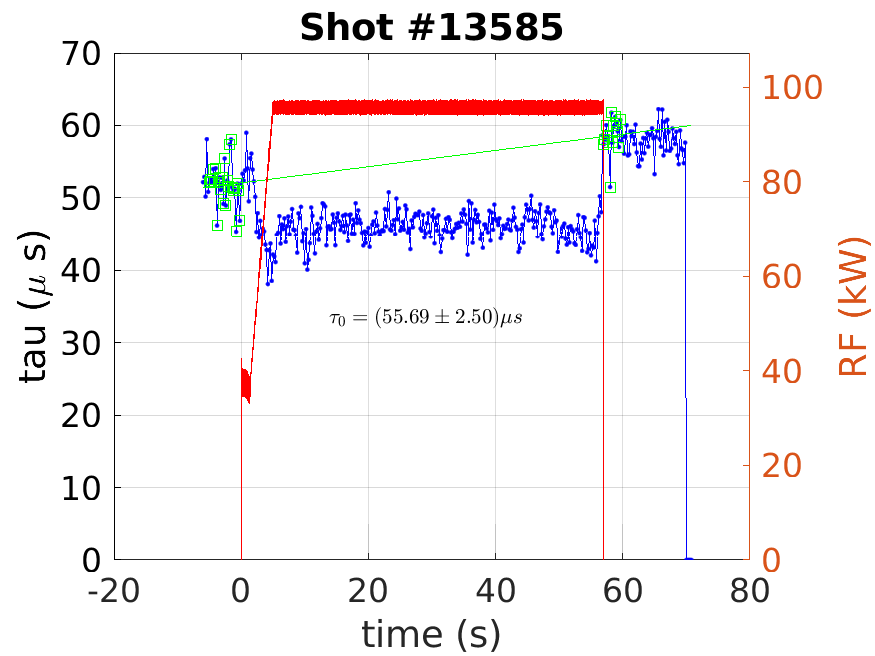}
        \caption{}
    \end{subfigure}
 \begin{subfigure}[b]{0.32\textwidth}
        \centering
        \includegraphics[width=\textwidth]{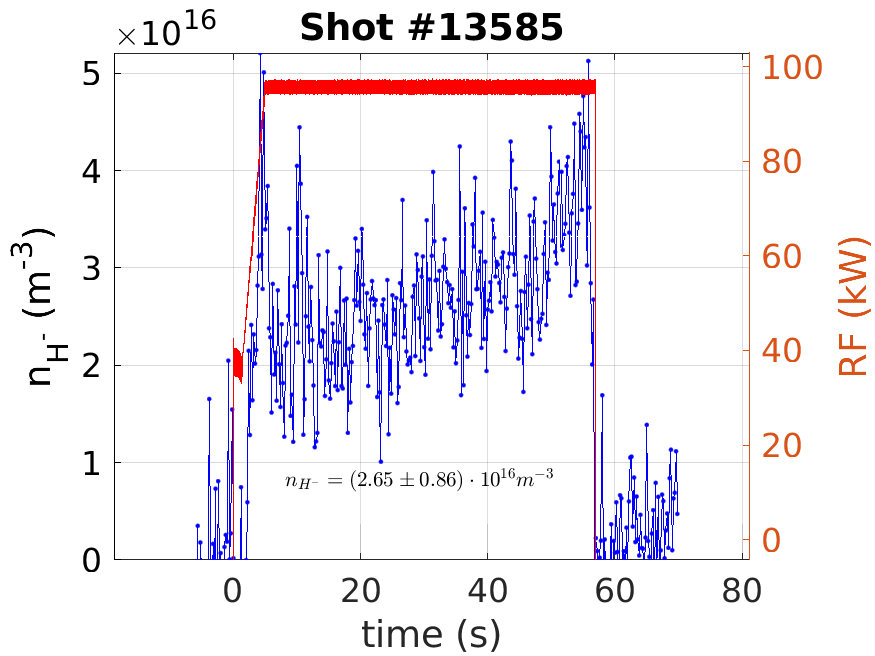}
        \caption{}
    \end{subfigure}
    \caption{(a) Single Ring-Down signal showing an exponential decay; (b) Decay time as a function of time during a plasma discharge with flat RF power; (c) Calculated H- density waveform. Source discharge parameters were RF power 100 kW, filter field 1.1 kA, source pressure 0.35 Pa.}
    \label{fig:main}
\end{figure}

The light exiting the cavity for each laser pulse exhibits an exponential decay, as shown in Fig.\,\ref{fig:LAS_CRDS_laser_path}(a). The signal is fitted with an exponential function to extract the characteristic decay time, $\tau$. The first 40\,$\mu$s of the signal are excluded from the fit to allow non-Gaussian modes to decay. When the plasma is on, the presence of H$^-$ ions leads to partial photodetachment, resulting in a faster absorption of the light stored within the optical cavity and thus in a shorter decay of the exiting light. Fig.\,\ref{fig:LAS_CRDS_laser_path} (b) shows the evolution of $\tau$ during a plasma pulse. The green points, recorded a few seconds before and just after the plasma discharge, are used as reference to calculate the baseline $\tau_0$ using a linear fit between the pre and post sets. 
Fig.\,\ref{fig:LAS_CRDS_laser_path} (c) shows the evolution of the evaluated negative ion density. Further details on this technique in SPIDER can be found in other references \cite{barbisan2021_CRDS}.

\subsubsection{Correlation with plasma emission spectroscopy}

\begin{figure}
	\centering
	\includegraphics[width=.8\columnwidth]{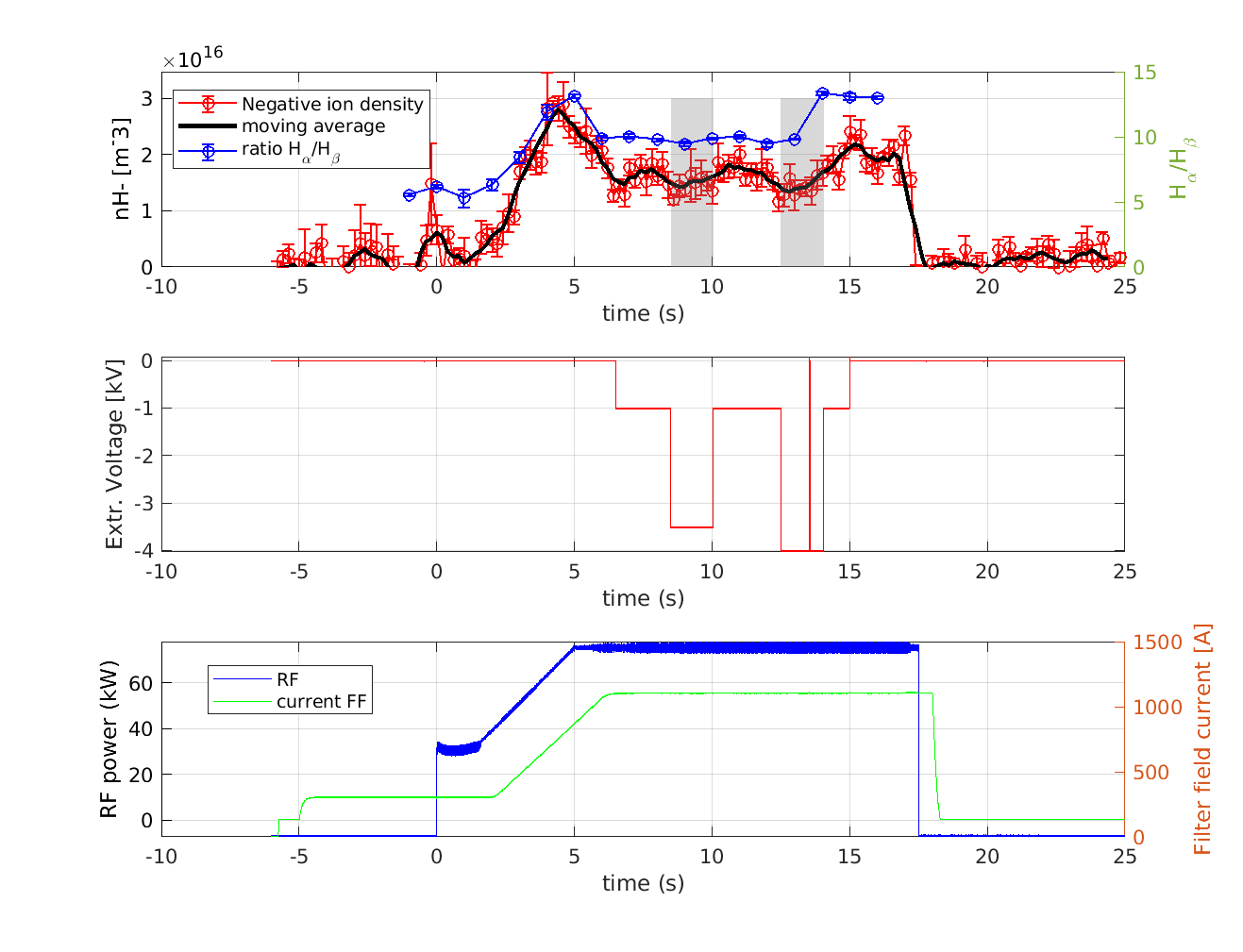}
	\caption{Top: $\mathrm{H^{-}}$ density and ratio of emission lines intensity $H_{\alpha}/H_{\beta}$, the shaded areas correspond to the beam extraction phases; center: RF power and filter field current; bottom: extraction voltage}
	\label{fig:CRDS_OES_trend}
\end{figure}

Due to the presence of drifts in the CRDS optical cavity, it is important to ensure that the variations in the decay time are actually linked to changes of negative ions. One possible approach is to compare the results with hydrogen Balmer line $H_{\alpha}$ emission measured through optical emission spectroscopy. In fact, it has been proposed in the past that, in negative ion sources such as SPIDER, the line ratio method can be used to infer information about the negative ion density \cite{fantz2006,ikeda2015}.
Indeed, close to the extraction region, where we have a cold recombining plasma and rich in negative ions, the recombination reaction is the dominant process:
\begin{equation}
    H^{+} + H^{-} \rightarrow H_{\alpha}
\end{equation}
In the recent experimental campaigns, a global decrease in the detected negative ion density compared to previous experiments with isolated beamlets has been observed. This reduction may be partly attributed to the new full-beam extraction conditions, since the opened apertures provide less surface area for negative ion production \cite{sartori2023}. 
Due to the limited signal-to-noise ratio, a sequence of four identical plasma pulses performed under the same conditions was selected for analysis. Only a very short sequence is taken to avoid variation of the plasma properties due to Cs conditioning.
Fig.\,\ref{fig:CRDS_OES_trend} on the top, shows the trend of the negative ion density and the ratio of the line emission $H_{\alpha}/H_{\beta}$ during a plasma pulse measured along a LOS between the PG and the BP, similar to that of CRDS and located just a few centimeters away.
Fig.\,\ref{fig:CRDS_OES_trend}, at the center, shows the extraction voltage.  Fig.\,\ref{fig:CRDS_OES_trend} at the bottom, shows the RF power with the intensity of the magnetic filter field. 
Both $n_{H^{-}}$ and the ratio $H_{\alpha}/H_{\beta}$ show two peaks along the execution of the plasma discharge, the first one at about $t=4\,s$ is mostly due to the variation of the magnetic filter field, which has an effect on the electron temperature, this trend had already been observed in SPIDER, but only in Cs-free conditions \cite{barbisan2022_Csfree}. Here we have confirmed this trend also in the presence of Cs. The second peak at the end of plasma discharge is due to the extraction voltage turning off at $t=15\,$s, therefore making more negative ions available in the extraction region. During beam accelerations, at $t=8.5\,$s and $t=12.5\,$s a decrease of $n_{H^-}$ is barely observable.
The main result from this observation is that  the trend of the line ratio can be used as an indicator of the variation of $H^{-}$ density during the plasma pulse. 
Since the two diagnostics (OES and CRDS) rely on different principles and are affected by different sources of uncertainty, their agreement reinforces the reliability of the measurements.

\subsubsection{CRDS upgrade}

\begin{figure}
	\centering
	\includegraphics[width=.80\columnwidth]{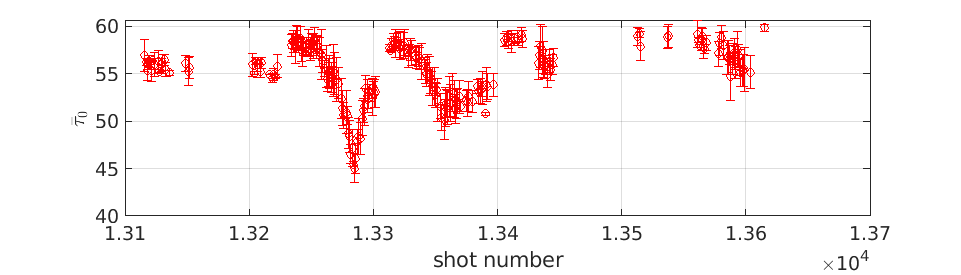}
	\caption{Trend of the CRDS cavity decay time throughout the experimental campaign before and after each plasma discharge.} 
	\label{fig:tau_0_history}
\end{figure}

Fig.\,\ref{fig:tau_0_history} shows the average decay time $\bar{\tau}_0$ of the cavity measured during several hundreds of shots before and after the plasma pulse. We remark that even though there are significant decreasing phases of $\bar{\tau}_0$, it approximately recovers to initial values. 
It is not clear whether the cause of these periodic variations is related to the mirrors’ reflectivity or to an issue with the optical alignment. In the case of SPIDER, the mirrors are located 1.830 m from the plasma box aperture, making it unlikely that the plasma could directly affect their reflectivity, even though an effect of UV radiation cannot be excluded.  Conversely, in ELISE, the mirrors are directly exposed to the plasma \cite{mimo2020} and may be influenced by Cs deposition, hot gas or UV radiation. 
A more plausible explanation is a periodic drift in the alignment, which is amplified by the long length of the optical cavity. Nevertheless, the influence of the internal vessel environment cannot be completely excluded. In fact, the mirrors used in a previous experimental campaign showed a complete loss of reflectivity and visible degradation of the reflective coating, most likely due to a significant water leak in the vessel at the end of that campaign. To mitigate such issues, the current mirrors are equipped with gate valves that remain closed when no experiments are running.
Overall, one of the critical challenges of this diagnostic system is maintaining the robustness of the laser alignment across multiple experimental campaigns.
\begin{figure}
	\centering
	\includegraphics[width=.55\columnwidth]{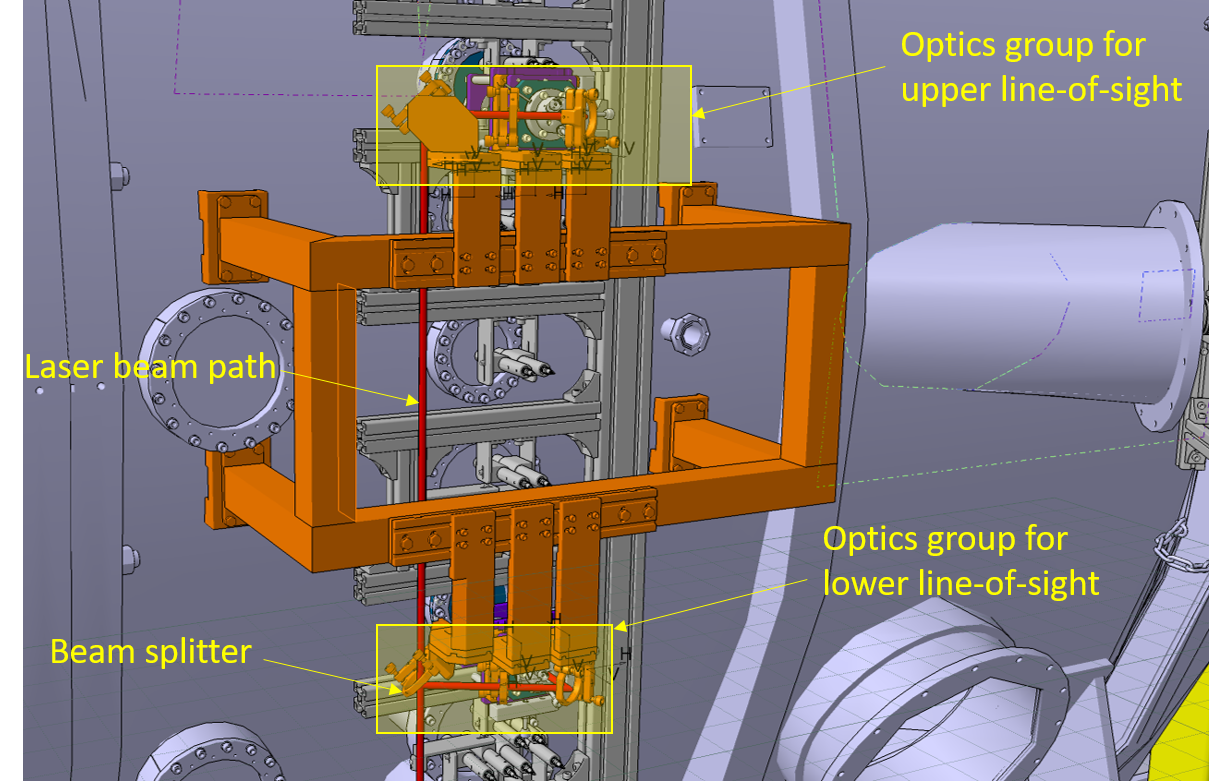}
	\caption{Development of a new structural design for the CRDS diagnostic aimed at enhancing robustness and incorporating a second line-of-sight positioned above the second beam segment.}
	\label{fig:upgrade_CRDS}
\end{figure}
On the basis of the experience of the previous years, it was realized that the alignment of CRDS is a very delicate process, this is mainly because of the very long laser path outside the cavity and the 4.637 m of the cavity itself. Also, the CRDS cavity components and input optics are mounted on the main structure together with LAS and OES. We have therefore decided to mechanically decouple the CRDS from other diagnostics by projecting a dedicated structure, which is highlighted in orange in the CAD design in Fig.\,\ref{fig:upgrade_CRDS}. 
Compared to the previous optical frame, which was anchored at two points on the vessel, the new frame is anchored at four points and reinforced with additional beams, thereby reducing the overall bending of the structure.
Since the robustness will be increased, a second line of sight, located at the symmetrical upper part of the vessel will be added.  The new frame will be anchored directly onto the vessel on welded plates. On the outer beams of the frame, extensions will be mounted to support the optical mounts for each line of sight. The laser beam, coming from the bottom, will be split at the level of the bottom mirror holder. A possible way to simplify laser injection into the cavity is to use an optical fiber. However, the beam focusing required on the optical component would greatly increase the laser power density, making thermal load management highly challenging. This would therefore require a dedicated feasibility study.

\subsection{Laser Absorption Spectroscopy}
The LAS diagnostic employs a Distributed Feedback (DFB) diode laser emitting near 852 nm. The emission wavelength is finely tuned by varying the drive current, with a tuning sensitivity of approximately 3 pm/mA. The current is sawtooth modulated to scan around the Cs D2 line transition $6^2P_{3/2} - 6^2S_{1/2}$ and its power is kept below $1\,W/m^{2}$ to avoid depopulation effects \cite{barbisan2019_LAS}.
\begin{figure}
	\centering
	\includegraphics[width=.4\columnwidth]{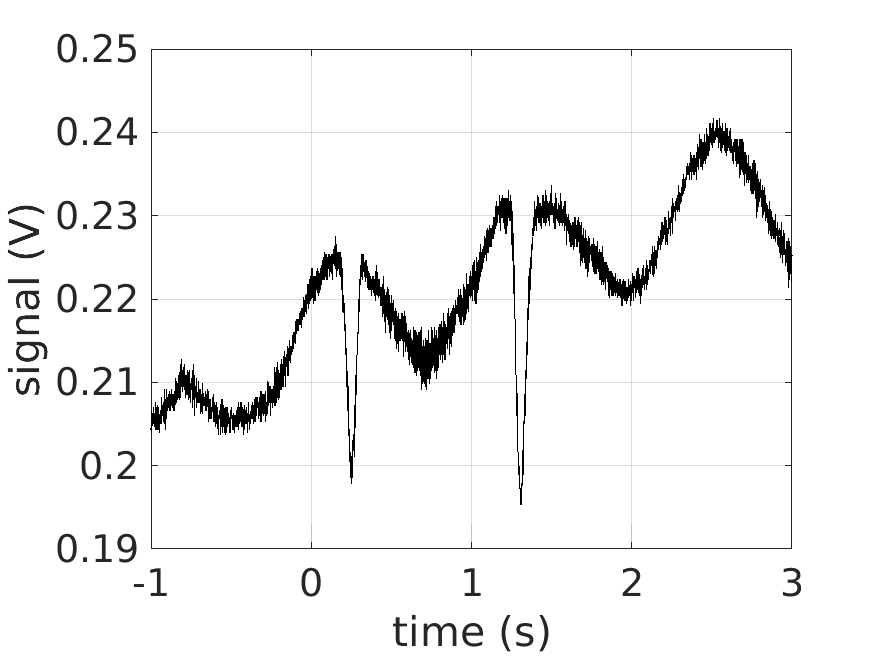}
	\caption{Transmitted LAS laser intensity signal as a function of time during a laser current ramp.}
	\label{fig:LAS_signal}
\end{figure}
Fig.\,\ref{fig:LAS_signal} shows an example of a wavelength scan with a current modulation frequency of 0.3 Hz. Unfortunately, due to interferences likely due to the vacuum windows, large oscillations are present in the background signal. Furthermore, higher sweep frequencies caused distortions of the baseline signal at the edges of the frequency ramp. These distortions arise from the active temperature regulation of the laser diode required during the frequency sweep. The horizontal wavelength scale is reconstructed from the known separation of the peaks (of 21.4 pm). The Cs density (at the ground state) can be calculated as follows:
\begin{equation}
    n_{Cs} = \frac{8 \pi c}{\lambda_0^4} \frac{g_k}{g_i} \frac{1}{A_{ik}l} \int \ln{\frac{I(\lambda,0)}{I(\lambda,l)}} d\lambda
\end{equation}
where $c$ is the speed of light, $\lambda_0 = 852.11$ nm is the D2 line wavelength, $g_k=2$ and $g_i=4$ are the statistical weights of lower and upper levels, $A_{ik} = 3.276 \cdot 10^7 \, Hz$ is the transition probability for the D2 line spontaneous emission, $l=0.87$ m is the effective LOS and $I(\lambda,0)$ and $I(\lambda,l)$ are the intensity of the laser beam before and after passing through the plasma.
Further details on this technique in SPIDER can be found on other references \cite{fantz2011_2,barbisan2019_LAS,barbisan2023_SOFT}. \\
In general, regarding the usefulness of the diagnostic for the experimental operation, the measurement of Cs density by LAS over time allows one to assess the stability of the conditioning achieved. For a given evaporation rate, the Cs density progressively increases with time (even within the same day) if the source is not yet conditioned. Conversely, once a stable condition is reached, this increase does not occur, and the same behavior is consistently reflected in the beam parameters.

\subsubsection{Cs in vacuum and plasma phases}

\begin{figure}[htbp]
    \centering
    \begin{subfigure}[b]{0.45\textwidth}
        \centering
        \includegraphics[width=\textwidth]{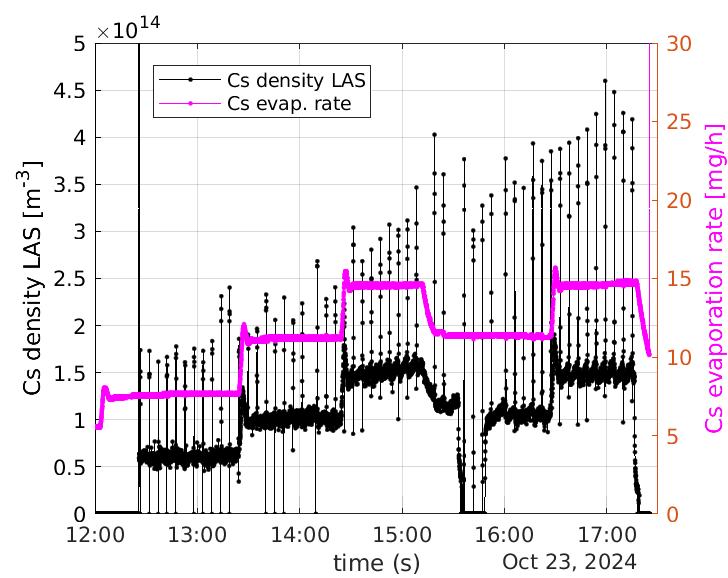}
        \caption{}
    \end{subfigure}
    \hfill
    \begin{subfigure}[b]{0.45\textwidth}
        \centering
        \includegraphics[width=\textwidth]{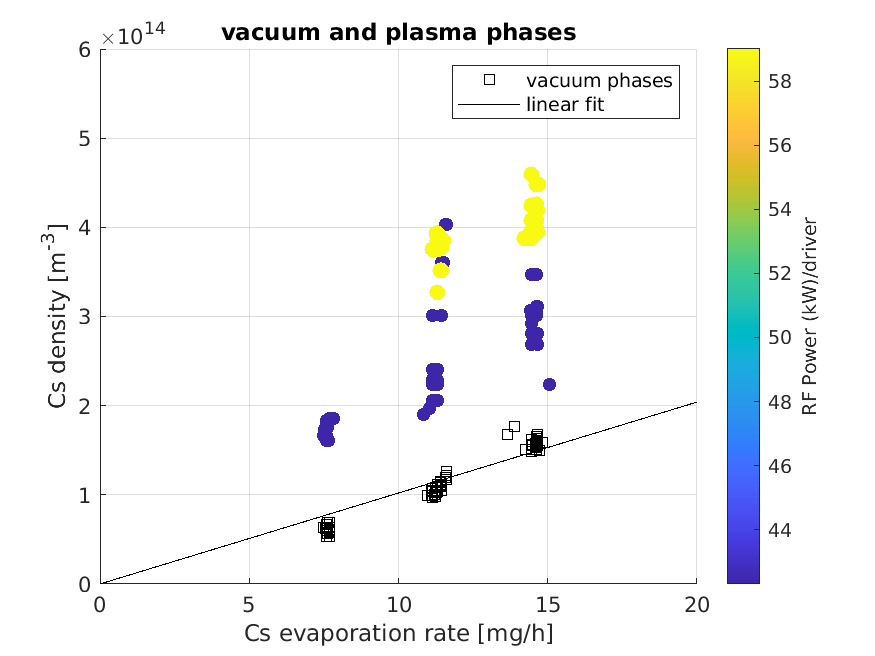}
        \caption{}
    \end{subfigure}
    \caption{(a) Cs density measured by LAS and evaporation rate during an experimental day; (b) Corresponding Cs density during vacuum (black squares) and plasma phases and effect of RF power.}
    \label{fig:LAS_day}
\end{figure}

In order to thoroughly understand the dynamics of Cs behavior, it is essential to know the condition of the wall surfaces, particularly the level of Cs coverage. For this purpose, a work function diagnostic would be ideal. This diagnostic is under development at the Cs facility CATS \cite{sartori2018} as a testbench for a following implementation in SPIDER.
Indeed, the characteristic of the Cs coverage strongly influences the production of negative ions and, consequently, the extracted current. In SPIDER, we currently rely on LAS for Cs diagnostics, which only provides the neutral Cs density in the source volume. 
Unfortunately, some studies in SPIDER attempting to correlate the extracted ion current with LAS measurements have not revealed any clear correlation and are therefore not shown here, however, in another plasma source, results suggest the existence of an optimum Cs density that maximizes the ion current \cite{fantz2012}.
The dynamics of Cs inside a negative ion source is indeed an interplay of many effects such as the oven evaporation directionality, the walls' conditions and the Cs redistribution effect due to the plasma.\\ 
One of the main topics of investigation regarding Cs dynamics is the effect of the Cs accumulation in the source volume due to oven evaporation, and that of the Cs redistribution due to plasma.
Figure\,\ref{fig:LAS_day} (a) shows the Cs density measured by LAS in black during an experimental day. In this case, regular, flat, and extended evaporation intervals allow for more reliable interpretation of the data.
Figure\,\ref{fig:LAS_day} (b) shows the corresponding Cs density as a function of the Cs evaporation rate, both during vacuum phases (black square markers) and plasma phases (colored circle markers, where the color indicates the RF power level). The Cs density during the vacuum phases is evaluated 120 seconds before the plasma pulse, while for the plasma phases, the maximum Cs density observed during the pulse is considered.
During vacuum phases, the increase in Cs density with respect to the total evaporation rate is ${\Delta\,{n_{Cs}}}/{ \Delta\,R_{ev}} = (1.02 \pm  0.01) \times 10^{13}\,{{m^{-3}} } {{[mg/h]^{-1}}}$, as calculated by the slope of the linear fit. This value is higher compared to the past of $0.451\,\cdot 10^{13} \,{{m^{-3}} } {{[mg/h]^{-1}}}$ measured in the 2021 experimental campaign \cite{sartori2022} probably because Cs was evaporated from all three ovens and thus distributed over a larger area, whereas in the recent campaign, Cs evaporation was provided by a single oven, located at the bottom of the source. 
During plasma phases, the increase in Cs density per unit of injected RF power, regardless of the evaporation rate, is estimated as: ${\Delta\,{n_{Cs}}}/{ \Delta\,P_{RF}} = (3.99 \pm  0.19) \cdot 10^{12}\,{{m^{-3}} } {{kW^{-1}}}$.
It should also be noted that when evaluating the increase in LAS-measured Cs density due to RF power, the result is valid only for that specific day. This means that repeating the same Cs evaporation scan with the same RF power on a later day (possibly weeks later) may yield a different response in Cs density, since the source conditions (e.g., wall Cs coverage, temperature, residual Cs, etc.) will have changed. The process of caesium sputtering and redistribution is therefore a process that also depend on the "history" of the source evaporation and operation.

During the last experimental campaign we also started to test relatively longer pulses, up to 2 min. This was the occasion not only to study the beam stability but also how the Cs dynamics is affected during a longer discharge.
Although the acquisition rate of LAS is only 0.3 Hz, is possible to identify the fast transient at the start and end of the plasma discharge.

\begin{figure}[htbp]
       \centering
        \includegraphics[width=6cm]{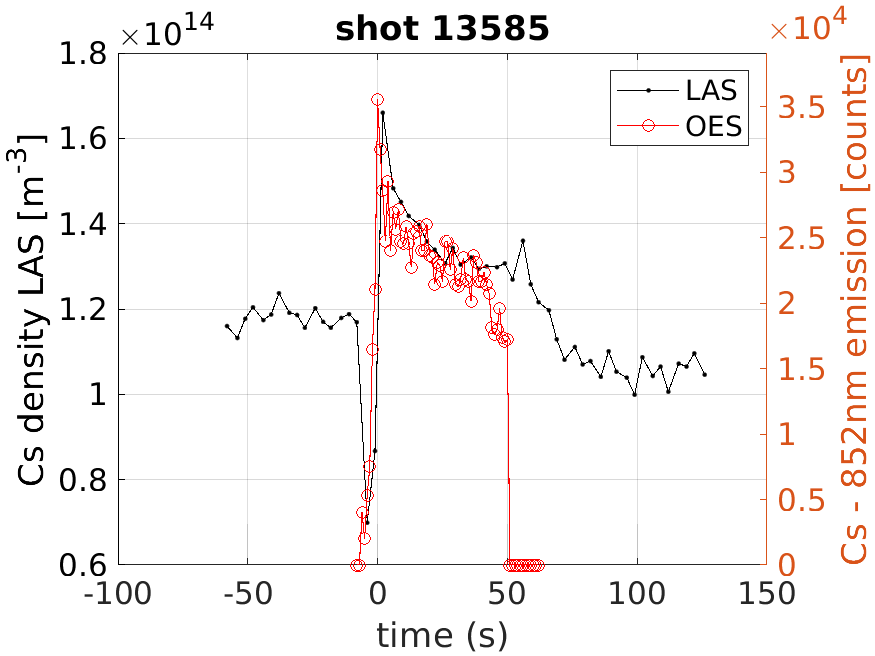}
    \caption{Left: Cs density measured by LAS at constant evaporation rate of 24 mg/h with 100 kW RF pulses; Right: Zoom on the plasma pulse and intensity of 852 nm recombination emission line.}
    \label{fig:LAS_and_OES}
\end{figure}

To confirm the drop of $n_{Cs}$ during plasma pulses, Fig.\, \ref{fig:LAS_and_OES}  shows, along with the $n_{\text{Cs}}$ values from LAS, the intensity of the characteristic Cs emission line at 852 nm, which corresponds to the $6^{2}P_{3/2} - 6^{2}S_{1/2}$ transition, for a LOS located 88 mm over the LAS LOS. It should be noted that, although the emission line detected by spectroscopy occurs at the same wavelength used for Cs laser absorption measurements, the spontaneous plasma emission at 852 nm is several orders of magnitude weaker than the probing laser and can thus be neglected in the Cs density estimate.
The emission intensity measured by optical emission spectroscopy shows a similar decreasing trend during the plasma pulse, supporting the observations from LAS. This decrease is also expected by simulations \cite{mimo2017}.
At the same time, to confirm that the reduction of the 852 nm line intensity is due to a decrease in Cs density (and not to a change in Cs excitation by the plasma), we verified plasma stability by monitoring the $H_{\alpha}$ line, which remains essentially constant during the plasma discharge.
At the end of the plasma discharge a rapid increase of $n_{Cs}$ due to Cs${^+}$ recombination with electrons is observed, as indicated by the data point at $t=56\,s$ in Fig.\,\ref{fig:LAS_and_OES}.\\ 
It has to be pointed out, however, that to quantitatively investigate the Cs dynamics, a full 3D model for Cs transport would be required, which is currently under development for SPIDER. This is justified by the fact that Cs${^+}$ transport across the plasma, which is also influenced by the magnetic field topology, also plays a role.

\section{Conclusion}
We have presented an overview of the laser diagnostics currently implemented in a full-scale negative ion source for fusion applications and reported recent results from the latest experimental campaign. These represent the first measurements obtained in the full-scale ion source prototype for ITER NBIs under conditions approaching nearly full extraction parameters.
Qualitative trends of both negative ion and Cs densities during plasma pulses have been shown and compared with optical emission spectroscopy measurements. We have shown that although CRDS has been routinely used in recent years, technical upgrades are still required to improve its reliability and robustness. In particular, the implementation of a second line of sight makes the mechanical upgrade even more crucial.
We have also presented data on the Cs density evolution during vacuum phases and the effect of RF power, a crucial aspect in the development of these sources. The data collected using these techniques provide a complementary tool to optimize these sources. 
Current efforts are focused both on improving the reliability of the experimental setups and on developing simulation tools capable of predicting Cs dynamics and negative ion production in the source.

\acknowledgments

This work has been carried out within the framework of the EUROfusion Consortium, partially funded by the European Union via the Euratom Research and Training Programme (Grant Agreement No 101052200 — EUROfusion). This work was carried out within the framework of the ITER-RFX Neutral Beam Testing Facility (NBTF) Agreement and has received funding from the ITER Organisation. The Swiss contribution to this work has been funded by the Swiss State Secretariat for Education, Research and Innovation (SERI). Views and opinions expressed are however those of the author(s) only and do not necessarily reflect those of the European Union, the European Commission or SERI. Neither the European Union nor the European Commission nor SERI can be held responsible for them.


\bibliographystyle{unsrt}
\bibliography{biblio}

\begin{thebibliography}{10}

\bibitem{hemsworth2017}
R.~S. Hemsworth et~al.
\newblock Overview of the design of the {ITER} heating neutral beam injectors.
\newblock {\em New J. Phys}, 19:025005, 2017.

\bibitem{marcuzzi2023}
D.~Marcuzzi et~al.
\newblock {L}essons learned after three years of spider operation and the first {MITICA} integrated tests.
\newblock {\em Fus. Eng. Des.}, 191:113590, 2017.

\bibitem{pasqualotto2012}
R.~Pasqualotto et~al.
\newblock {D}esign of laser-aided diagnostics for the negative hydrogen ion source {SPIDER}.
\newblock {\em JINST}, 7:C04016, 2012.

\bibitem{quandt1999}
E.~Quandt et~al.
\newblock {M}easurement of negative-ion densities by cavity ringdown spectroscopy.
\newblock {\em Europhys. Lett.}, 45:pp. 32--37, 1999.

\bibitem{tsumori2021}
K.~Tsumori et~al.
\newblock A review of diagnostic techniques for high-intensity negative ion sources.
\newblock {\em Appl. Phys. Rev.}, 8:021314, 2021.

\bibitem{fantz2011}
U.~Fantz et~al.
\newblock Quantification of {C}esium in negative hydrogen ion sources by laser absorption spectroscopy.
\newblock {\em AIP Conf. Proc.}, 1390:348–358, 2011.

\bibitem{heinemann}
B.~Heinemann et~al.
\newblock {T}he negative ion source test facility {ELISE}.
\newblock {\em Fus. Eng. Des.}, 86:768--771, 2011.

\bibitem{fantz2006}
U.~Fantz et~al.
\newblock A novel diagnostic technique for {H-} ({D}-) densities in negative hydrogen ion sources.
\newblock {\em New Journal of Physics}, 8:301, 2006.

\bibitem{poggi2023}
C.~Poggi et~al.
\newblock {M}easure of negative ion density in a large negative ion source using {L}angmuir probes.
\newblock {\em Journal of Instrumentation}, 18:C08013, 2023.

\bibitem{zagorski2025}
R.~Zagorski et~al.
\newblock Numerical reconstruction of {L}angmuir probe measurements obtained from the negative ion source for {ITER} {(SPIDER)}.
\newblock {\em Plasma Phys. Control Fusion}, 67:065020, 2025.

\bibitem{pouradier2024}
B.~Pouradier-Duteil et~al.
\newblock {C}haracterization of plasmas in negative ion sources using a {C}s-{H} collisional radiative model.
\newblock {\em Journal of Instrumentation}, 19:C02051, 2024.

\bibitem{barbisan2022_Csfree}
M.~Barbisan et~al.
\newblock {N}egative ion density in the ion source {SPIDER} in {C}s free conditions.
\newblock {\em Plasma Phys. contol. Fusion}, 64:065004, 2022.

\bibitem{barbisan2021_CRDS}
M.~Barbisan et~al.
\newblock {D}evelopment and first operation of a cavity ring down spectroscopy diagnostic in the negative ion source spider.
\newblock {\em Rev. Sci. Instrum.}, 92:053507, 2017.

\bibitem{barbisan2023_SOFT}
M.~Barbisan et~al.
\newblock Characterization of cesium and ${H^-}/{D^-}$ density in the negative ion source {SPIDER}.
\newblock {\em Fus. Eng. Des.}, 194:113923, 2023.

\bibitem{sartori2023}
E.~Sartori et~al.
\newblock Influence of plasma grid-masking on the results of early {SPIDER} operation.
\newblock {\em Fus. Eng. Des.}, 194:113730, 2023.

\bibitem{barnett1977}
Barnett~C. F. et~al.
\newblock {A}tomic {D}ata for {C}ontrolled {F}usion {R}esearch.
\newblock Technical Report Vol. 1, ORNL, February 1977.

\bibitem{ikeda2015}
K.~Ikeda et~al.
\newblock Evaluation of negative ion distribution changes by image processing diagnostic.
\newblock {\em AIP Conf. Proc.}, 1655:040005, 2015.

\bibitem{mimo2020}
A.~Mimo et~al.
\newblock {C}avity ring-down spectroscopy system for evaluation of negative hydrogen ion density at the {ELISE} test facility.
\newblock {\em Rev. Sci. Instrum.}, 91:013510, 2020.

\bibitem{barbisan2019_LAS}
M.~Barbisan et~al.
\newblock Design and preliminary operation of a laser absorption diagnostic for the {SPIDER} {RF} source.
\newblock {\em Fus. Eng. Des.}, 146:2707–2711, 2019.

\bibitem{fantz2011_2}
U.~Fantz et~al.
\newblock {O}ptimizing the laser absorption technique for quantification.
\newblock {\em J. Ohys. D: Appl. Phys.}, 44:335202, 2011.

\bibitem{sartori2018}
E.~Sartori et~al.
\newblock Diagnostics of caesium emission from {SPIDER} caesium oven prototype.
\newblock {\em AIP Conf. Proc.}, 2052:040011, 2020.

\bibitem{fantz2012}
U.~Fantz et~al.
\newblock Cesium dynamics in long pulse operaiton of negative hydrogen ion sources for fusion.
\newblock {\em Rev. Sci. Instrum.}, 83:02B110, 2012.

\bibitem{sartori2022}
E.~Sartori et~al.
\newblock First operation with caesium of the negative ion source {SPIDER}.
\newblock {\em Nucl. Fusion}, 62:086022, 2022.

\bibitem{mimo2017}
A.~Mimo et~al.
\newblock Modelling of {C}aesium {D}ynamics in the {N}egative {I}on {S}ource at {BATMAN} and {ELISE}.
\newblock {\em {AIP} Conf. Proc.}, 1869:030019, 2017.

\end{thebibliography}

\end{document}